# On estimation of fractal dimension for 2D time-series based on functional relation between areas of covers

Dmitry Zhabin[1]
Integrated Risk Management Department
VTB (PJSC), Moscow, 123112, Presnenskaya emb., 12

It is shown that fractal dimension can be estimated seeking a solution of functional equation defined for areas of coverages of different scales. The method proposed is compared with widely known way to estimate fractal dimension via linear regression for numbers of ordinary sets, which are used to cover the fractal, and scale size. Due to its simplicity the method described in the article may be useful to get estimation of fractal dimension for 2D time-series.

## 1. Introduction.

The problem of fractal dimension estimation is well known while analyzing irregular objects. As a general definition of fractal dimension the Hausdorf dimension [1] may be used but it rather complicated in practical applications, so Minkowsky dimension [2] may be the best choice while seeking the way to get practical estimation of a fractal dimension (while for particular systems other dimension definitions that suits more for specific peculiarity may be used, for details see [3]).

Mkinkowsky dimension is usually estimated through linear regression, as described in [4] for example. This relatively simple way to get practical estimation needs nevertheless several points of observations for regression to be applied to, which means several scales of time-series to be considered.

[1] E-mail address: dzhabin@vtb.ru

In the present article we show the expression for the fractal dimension estimation, which requires two scales of time-series to get the estimation, and then compare this result with the one of linear regression application.

## 2. Fractal dimension estimation and its applications

According to [5], for fractal objects for which Minkowsky dimension exists, it is true that areas of covers for different scales ought to follow a functional relation:

$$S(\alpha\delta) = \alpha^{D_E - D_H} S(\delta) \,, \tag{1}$$

which means that knowing areas of covers for different scales one can rewrite Minkowsky dimension definition in a following way:

$$D_H = D_E - \lim_{\delta \to 0} \left[ \frac{ln\left(\frac{S(\alpha\delta)}{S(\delta)}\right)}{ln(\alpha)} \right]. \tag{2}$$

Here $S(\delta)$ – is the area of cover for the given size of the cover element δ; $S(\alpha\delta)$ – is the area of cover, when size of the cover element is scaled in α times; $D_E$ – is the dimension of Euclidian space in which the fractal embedded into; $D_H$ – is the fractal dimension of the given fractal object.

### 2.1 Application of expression (2) for Cantor set.

Fractal dimension of Cantor set is well known and equal to $\ln(2)/\ln(3)$, see [3]. The $n^{th}$ level of set hierarchy can be covered by intervals with overall length equal to $S(n) = 2^n/3^n$. Then the estimation of the fractal dimension with help of expression (2) takes form:

$$D_H = 1 - \frac{\ln\left(\frac{2^{n-1}}{3^{n-1}} \cdot \frac{3^n}{2^n}\right)}{\ln(3)} = \frac{\ln(2)}{\ln(3)}. \tag{3}$$

### 2.2 Fractal dimension estimation for Koh set.

Koch snowflake is another well-known fractal, which Hausdorff dimension is $\ln(4)/\ln(3)$, see [3]. For $n^{th}$ step of iteration procedure used for snowflake construction the area of the cover made by equal boxes with size $1/3^n$ is equal to $S(n) = 4^{n-1}/(3^n)^2$. So, the application of the expression (2) to Koch set gives fractal dimension as:

$$D_H = 2 - \frac{\ln\left(\frac{4^{n-2}}{3^{2(n-1)}} \cdot \frac{3^{2n}}{4^{n-1}}\right)}{\ln(3)} = \frac{\ln(4)}{\ln(3)}. \quad (4)$$

**2.3 Fractal dimension estimation for 2D time-series**

Expression (2) can be applied also to calculate proxy of fractal dimension for financial market quotes. Indeed, time series of financial markets prices are widely presented in terms of Japanese candlesticks [6], when each discrete period of time is parameterized by four numbers: open price (first price of the period); high price (maximal price observed during the period); low price (minimal price observed during the period); close price  (period last price). If open price is higher than close one, then the candlestick has white color or is black otherwise. Such parameterization perfectly fits for calculation of area needed to cover price walk.

*RTSI* is an index on Russian stock market quoted by Moscow exchange. Sample of the *RTSI* quotes can be downloaded from public sources: *www.finam.ru*. We use following time-series scales for the analysis: 1 minute (see Figure 1); 5 minutes;10 minutes; 15 minutes; 30 minutes; 1 hour and 9 hours (which is equal to 1-day time step, see Figure 2). We took fourth quarter of 2019 as a time-period of our investigation.

Firstly, let us proceed with standard way of fractal dimension calculation, when linear regression is applied to parametrize dependence between logarithms of numbers of ordinal sets and logarithms of scales:

$$\ln(N) = -D_f \cdot \ln(\delta) + C \quad (5)$$

Here $N$ – number of squares with side length δ; $D_f$ – fractal dimension; C – an arbitrary constant. For the specified timescales and given time-series we can easily get data for further regression application:

**Table 1.** Data for linear regression calculation based on the specified time-series sample.

| | 1min | 5min | 10min | 15min | 30min | 60min | 1d |
|---|---|---|---|---|---|---|---|
| δ | 0,00000190 | 0,00000951 | 0,00001903 | 0,00002854 | 0,00005708 | 0,00011416 | 0,00102740 |
| N | 8 467 353 270 | 883 275 011 | 327 061 953 | 183 137 986 | 68 859 767 | 25 349 603 | 1 057 341 |
| -ln(δ) | 13,17 | 11,56 | 10,87 | 10,46 | 9,77 | 9,08 | 6,88 |
| ln(N) | 22,86 | 20,60 | 19,61 | 19,03 | 18,05 | 17,05 | 13,87 |

The result of the linear regression is presented on the Figure 3 and it gives fractal dimension estimation equals to 1.4291.

Alternatively, we can get the fractal dimension estimation by applying (2) to the given time-series. Let's denote by $H_i^{1m}$ and $L_i^{1m}$ maximal and minimal prices for $i^{th}$ 1-minutes Japanese candlestick. $H_i^{1D}$ and $L_i^{1D}$ denote maximal and minimal prices for $i^{th}$ 1-day candlestick (here one need to remember that 1day equals to 9 hours technically). So, estimation of fractal dimension for specified *RTSI* time-series takes form:

$$\widehat{D}_H = 2 - \frac{1\, ln\left(4 \cdot \sum_{i=1}^{N_D}\left(H_i^{1D} - L_i^{1D}\right) / \sum_{i=1}^{N_{1m}}\left(H_i^{1m} - L_i^{1m}\right)\right)}{ln(\delta_D / \delta_{1m})} \cdot= 1.4286. \quad (6)$$

Here $N_D$ – is number of Japanese candlesticks with 1-day duration; $N_{1m}$ – number of 1-minute Japanese candlesticks.

## 4. Conclusions

Fractal dimension is commonly treated as a proper measure of the system complexity. Meanwhile, not for all cases the fractal dimension can be analytically calculated, so methods that propose simple way to get estimation for fractal dimension are of some interest.

In the present article new method to estimate fractal dimension is presented, which is based on a functional equation defined for areas of covers with different scales. To demonstrate applicability of the method fractal dimensions for Cantor set, Kock snowflake and RTSI index are estimated.

It worth to mention the simplicity of the method for practical use since it requires only two scales to get a proxy measure of fractal dimension, which makes calculation not so consuming as it would be in case of well-known estimation of fractal dimension via linear regression routine.

**References.**

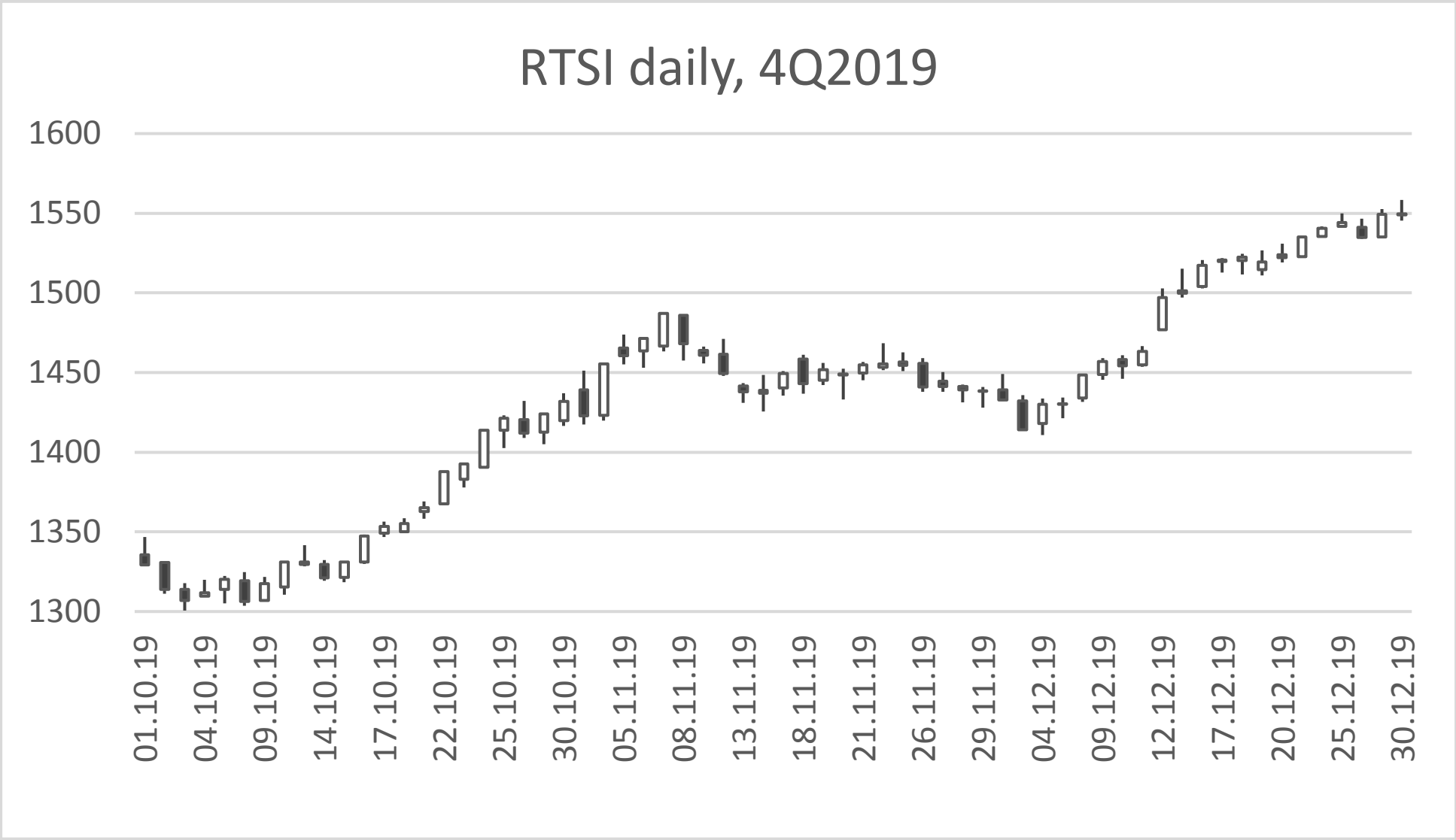

**Picture 1.** RTSI quotes in 4Q 2020. 1 day candle-sticks.

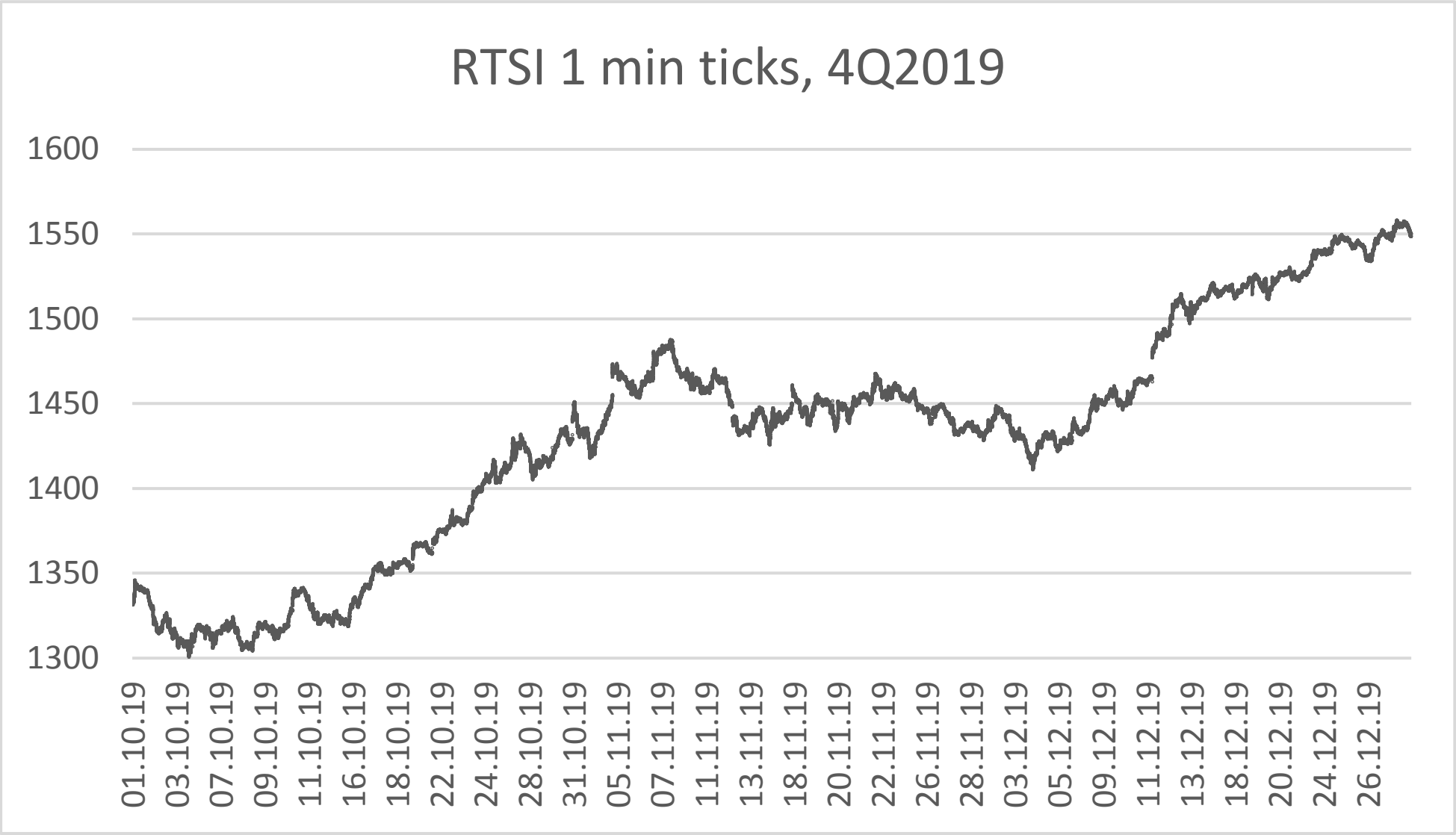

**Picture 2.** RTSI quotes in 4Q 2020. 1 minute candle-sticks.

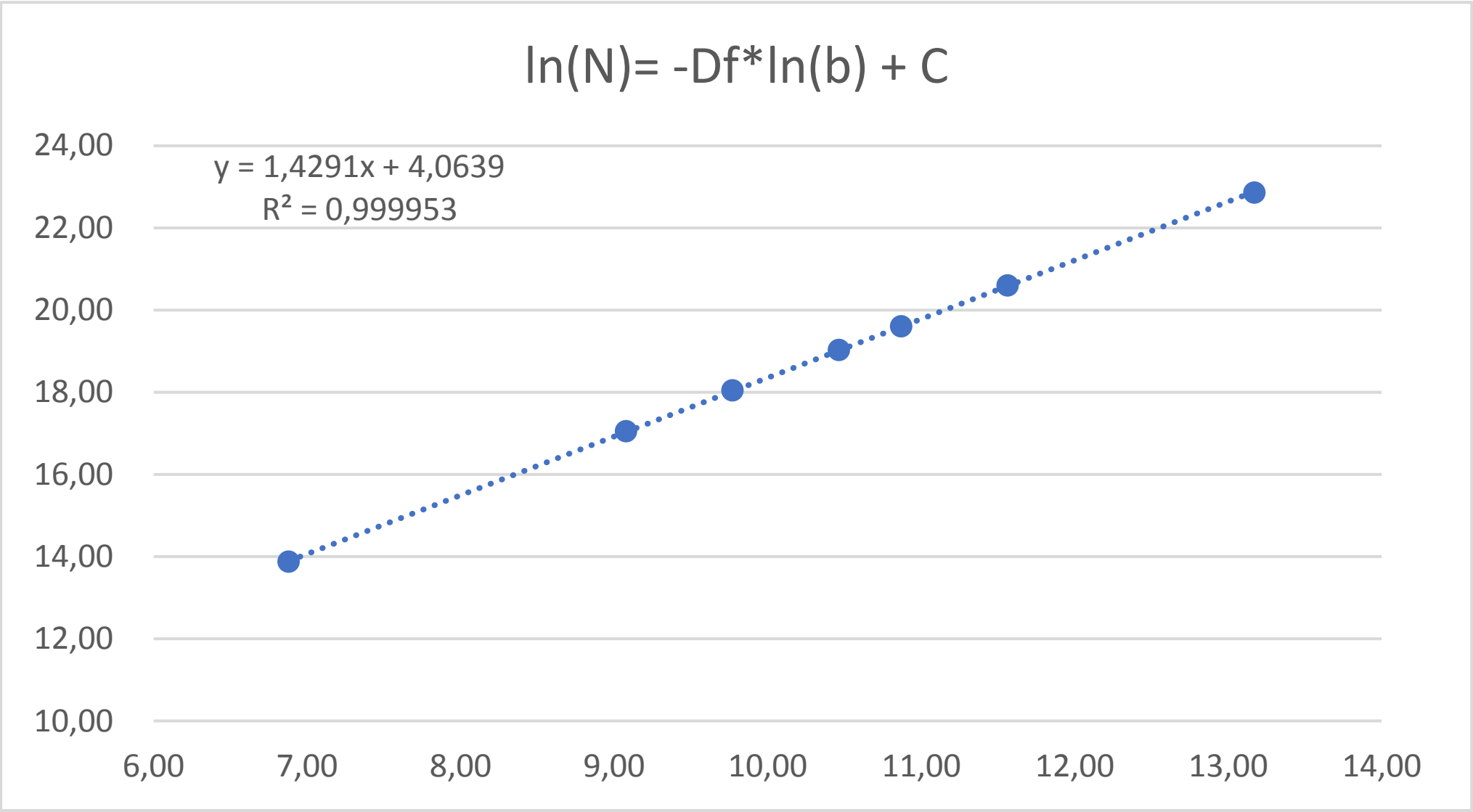

**Picture 3.** Linear regression applied for fractal dimension estimation.